\documentclass[12pt]{iopart}

\usepackage[abbr]{harvard}
\usepackage[utf8]{inputenc}
\usepackage{graphicx}
\usepackage{array}
\usepackage[usenames, dvipsnames]{color}
\usepackage{textcomp}
\usepackage{multirow}
\usepackage{contour}
\usepackage{verbatim}
\usepackage{float}
\usepackage{tikz}
\usepackage{color}

\begin{document}

%\title[]{{\it In-vivo} measurement of sub-resolution flow velocity profiles in microvessels using contrast-enhanced super-resolution technique}

%\title[]{Quantitative {\it In-vivo} measurement of sub-resolution flow velocity profiles in microvessels using contrast-enhanced super-resolution technique}

\title[]{Quantitative sub-resolution blood velocity estimation using ultrasound localization microscopy {\it ex-vivo} and {\it in-vivo}.}

\author{David~Esp\'indola$^{a,b}$, Ryan~M.~DeRuiter$^a$, Francisco~Santibanez$^a$, Paul~A.~Dayton$^a$ and Gianmarco~Pinton$^a$}

\address{$^a$ Joint Department of Biomedical Engineering, University of North Carolina at Chapel Hill and the North Carolina State University, Chapel Hill, North Carolina, USA.}
\address{$^b$ Currently at Instituto de Ciencias de la Ingenier\'ia, Universidad de O'Higgins,\\ Avenida Libertador Bernardo O'Higgins 611, Rancagua, Chile.}
\ead{gia@email.unc.edu}
\vspace{10pt}
\begin{indented}
\item[]\today
\end{indented}

\begin{abstract}
Super-resolution ultrasound imaging relies on the sub-wavelength localization of microbubble contrast agents. By tracking individual microbubbles, the velocity and flow within microvessels can be estimated.  It has been shown that the average flow velocity, within a microvessel ranging from tens to hundreds of microns in diameter, can be measured. However, a 2D super-resolution image can only localize bubbles with sub-wavelength resolution in the imaging plane whereas the resolution in the elevation plane is limited by conventional beamwidth physics. Since ultrasound imaging integrates echoes over the elevation dimension, velocity estimates at a single location in the imaging plane include information throughout the imaging slice thickness. This slice thickness is typically a few orders or magnitude larger than the super-resolution limit. It is shown here that in order to estimate the velocity, a spatial integration over the elevation direction must be considered. This operation yields a multiplicative correction factor that compensates for the elevation integration. A correlation-based velocity estimation technique is then presented. Calibrated microtube phantom experiments are used to validate the proposed velocity estimation method and the proposed elevation integration correction factor. It is shown that velocity measurements are in excellent agreement with theoretical predictions within the  considered range of flow rates  (10 to 90 $\mu$L/min). Then, the proposed technique is applied to two {\it in-vivo} mouse tail experiments imaged with a low frequency human clinical transducer (ATL L7-4) with human clinical concentrations of microbubbles. In the first experiment, a vein was visible with a diameter of 140~$\mu$m and a peak flow velocity of 0.8~mm/s. In the second experiment, a vein was observed in the super-resolved image with a diameter of ~120~$\mu$m and with maximum local velocity of $\approx$~4.4~mm/s. 
It is shown that the parabolic flow profiles within these micro-vessels are resolvable.
\end{abstract}

\vspace{2pc}

\newpage
\begin{indented}
\item[]
\section*{Introduction}

Recently, ultrasound localization microscopy (ULM), has received a significant amount of attention due to its ability to super-resolve ($\lambda/20$~\cite{hingot2019microvascular}) the vasculature well beyond the diffraction limit ($\lambda/2$)~\cite{couture2011microbubble,desailly2013sonoactivated}. ULM was inspired by optical localization microscopy, in which the center of stochastically blinking fluorescent sources are localized with sub-resolution accuracy~\cite{rust2006subdiffraction}.  Super-resolved images can be generated by accumulating the positions of the source centers in hundreds or thousands of images. In optics, this technique has been used to image intracellular proteins~\cite{betzig2006imaging} and the adhesion of live cells~\cite{shroff2008live} at nanometric scales,  providing new understanding of cellular architecture ~\cite{creech2017superresolution}. A similar approach can be used in ultrasound, where gas-filled encapsulated microbubble (MB) contrast agents with a number weighted mean diameter on the order of 1-5~$\mu$m act as blinking sources. The MBs, which are injected in the bloodstream, flow within the circulatory system, where they appear and disappear from the imaging region, enabling the sub-resolution localization of the center of individual MBs. Then, the super-resolved image is constructed from the accumulation of thousands to hundreds of thousands localized MBs. However, unlike  optical super-resolution techniques, where the blinking contrast agents are stationary, ultrasound contrast agents flow within the bloodstream. Thus, in addition to generating   super-resolved images, ultrasound can be used to generate spatially super-resolved velocity maps of blood flow. An extensive review of the ultrasound localization microscopy techniques can be found at \cite{couture2018ultrasound}.

Spatiotemporal source separability is the single most important requirement to localize the center of microbubbles and to overcome the diffraction limit of resolution. This has motivated  approaches that use high contrast agent dilutions ($\approx$~1600~MBs/mL) which are a thousand times smaller than what is used in conventional contrast-ultrasound-imaging to facilitate localization~\cite{oreilly2013superresolution}. However, low bubble concentrations require long acquisition times ($>$10~min) to populate an image. This makes the imaging system highly susceptible to motion artifacts and limits the clinical translatability of this approach. Higher MBs concentration have been used to reduce the acquisition time to clinically relevant intervals ($<$10 min)~\cite{errico2015ultrafast}. This approach,  coupled with a translation system, has also been used to produce 3D images of tumor-associated angiogenesis~\cite{lin2016super,lin20173-D}. 
In addition to these challenges, a low mechanical index should be used to avoid MB destruction, which, constrains the penetration depth to a maximum of 2~cm, due to low signal to noise ratio arising from acoustic attenuation. This has motivated the development of beamforming strategies that increase the penetration depth, such as  focused emissions \cite{espindola2018adaptive}.

 To estimate spatially super-resolved velocity maps, individual MBs must be tracked while they are flowing with the bloodstream. The nearest neighbor algorithm is the simplest technique to obtain the velocity vector of a microbubble between two frames~\cite{errico2015ultrafast}. This technique has been used to determine the in-plane velocity map of the blood flow in an {\it in-vivo} mouse brain, with a dynamic range in flow between 3~mm/s to a few ~cm/s. Another approach that estimates  blood flow velocity map is based on the cross-correlation algorithm~\citeasnoun{christensen2015invivo}. The cross-correlation of the intensity of each MB was used to estimate the velocity vectors and the corresponding velocity profiles in the microvasculature of an {\it in-vivo} mouse ear. Although both of these techniques have produced realistic velocity estimates that are in the expected ranges, they have not been directly validated, either by direct measurement of the velocity or by theory.

Recently, another bubble tracking approach has been developed. This relies on linear flow models and a probabilistic method based on a Markov chain Monte Carlo data association algorithm has been proposed ~\cite{ackermann2016detection}. In this work, a resolution beyond the diffraction limit has been produced, the velocity profile in a single microtube phantom was estimated and compared with the theoretical predictions. In addition, Ackermann et al. have presented a correction method for the velocity estimates that operates in the elevational plane. However, the validation presented here for the velocity profile was done in a tube that is bigger than the diffraction limit. The size of the tube was 400~$\mu$m in diameter and the axial resolution was 40~$\mu$m. This technique was also demonstrated in-vivo for tumor characterization \cite{song2018Ontheeffect}.

In this paper, we propose a technique that uses cross-correlation applied to a beamformed radio frequency (RF) data to estimate velocity maps of blood flow within vessels smaller than the diffraction limit. It is shown that the proposed technique can image the parabolic velocity profiles in an {\it in-vivo} mouse tail vein of 140~$\mu$m with a diagnostic imaging probe designed for human clinical use which operates at a relatively low-frequency (5.2~MHz). First, our technique is validated by comparing theoretical predictions with the velocity estimates for a microtube phantom experiment. The effect of the elevation plane on the measurement profiles is calculated theoretically and applied to the velocity estimates to determine quantitative velocity estimates, similarly to the correction found in ~\cite{ackermann2016detection}. This elevation plane correction is based on the assumption that the flow is laminar and unidirectional inside of the microvasculature. These conditions are easily met in small vessels due to their small Reynolds numbers~\cite{fung1973stochastic,landau1959course}. Finally, we demonstrate the use of the proposed velocity estimation technique in two {\it in-vivo} mouse tail experiments. These two experiments serve as a demonstration at two different blood flow regimes. In the first, the low velocity regime is explored with a peak velocity of less than 1~mm/s. In the second, by exploring a different anatomical location, a peak velocity of 4~mm/s was found.

\section*{Methods}
Experimental ultrasound acquisitions were performed using a programmable Verasonics (Kirkland, WA, USA) Vantage ultrasound scanner driving a commercial ATL L7-4 128-element linear ultrasound array (Philips, Andover, MA, USA)  operating at a center frequency of 5.2~MHz. The ultrasound scanner was programmed to emit plane wave imaging sequences  with a pulse repetition frequency of 700~Hz, a pulse duration of 1.5 cycles, and a transmit pressure of 131~kPa (mechanical index of 0.058). This produces a conventional B-mode lateral resolution given by $1.22\lambda$F-number $\approx 363~\mu$m, and an axial resolution of $\approx 222~\mu$m. Plane wave sequences, which  are one of the most simple imaging sequences to program, have been used reliably by several groups to generate ultrasound localization images \cite{errico2015ultrafast,christensen2015invivo}.  Plane wave imaging sequences were therefore chosen for their robustness and ease of implementation (also see Section \ref{discussion}), however the proposed velocity estimation technique can be implemented for other non-planar super-resolution imaging sequences~\cite{espindola2018adaptive}.

Lipid-encapsulated microbubble contrast agents containing decafluorobutane (SynQuest Labs, Alachua, FL, USA) were utilized in described studies.  These microbubbles were prepared as described previously~\cite{shelton2016molecular}, resulting in a polydisperse size distribution of contrast agents. The microbubbles were characterized using an Accusizer 780A Autodiluter Single Particle Optical Sizing (SPOS) System (PSS Nicomp, Santa Barbara, California, USA), which is sensitive to particle diameters within the range of 0.5 and 2500 $\mu$m. The prepared microbubble stock had a number-weighted mean diameter of ~1$\mu$m and a concentration of $10^{10}$ bubbles per milliliter.

\begin{figure}[t]
\begin{flushright}
\setlength{\unitlength}{0.84\textwidth}
\begin{picture}(1,0.48)(0,0)
%%%% frame
%\multiput(0,0)(0, 0.48){2}{\line(1,0){1}}
%\multiput(0,0)(1, 0){2}{\line(0,1){0.48}}
%%%%%%%%%
\put(0.03,0){\includegraphics[trim= 370 20 340 100,clip,width=0.45\unitlength]{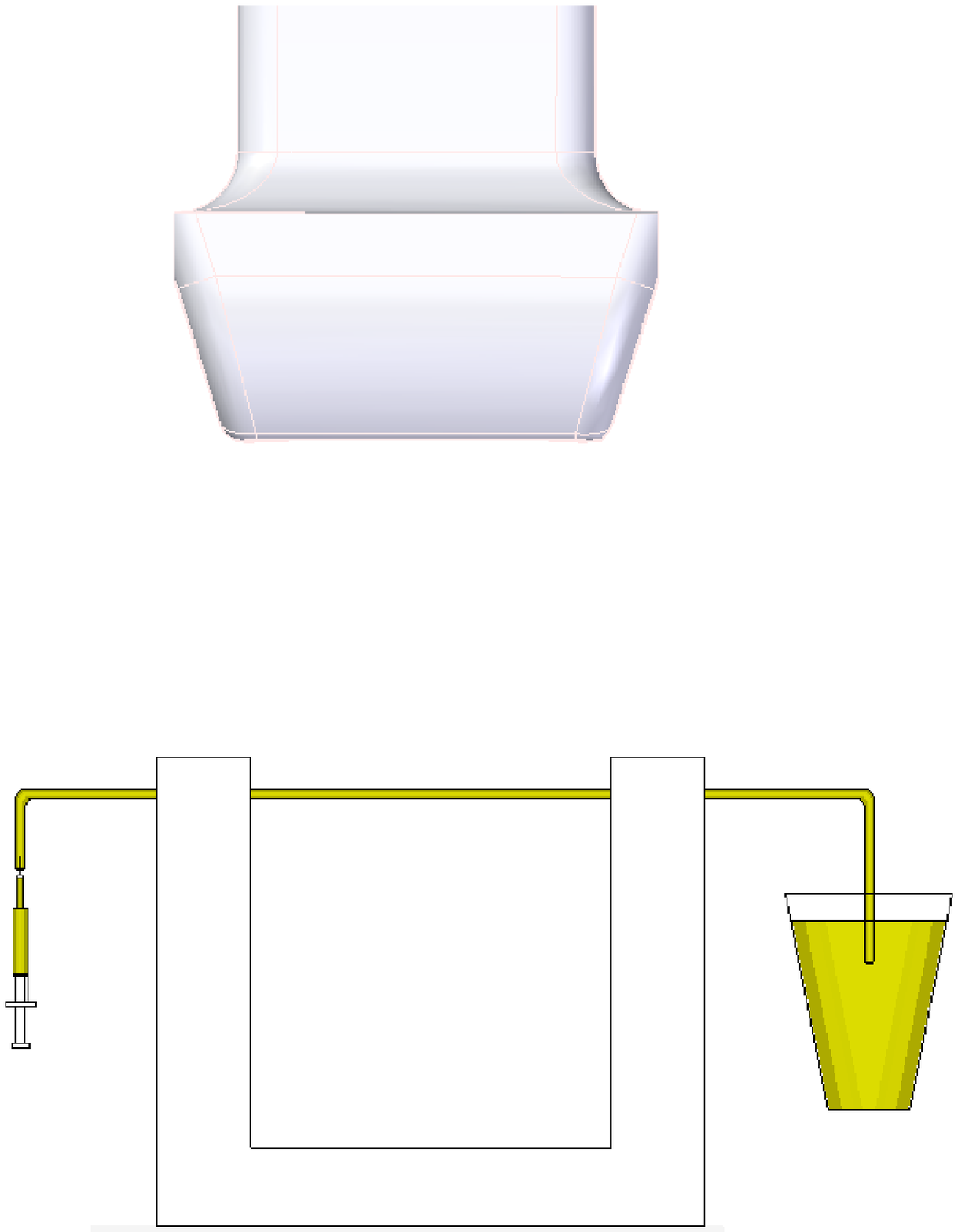}}
\put(0.53,0.22){\reflectbox{\includegraphics[trim= 365 55 130 100,clip,width=0.4\unitlength]{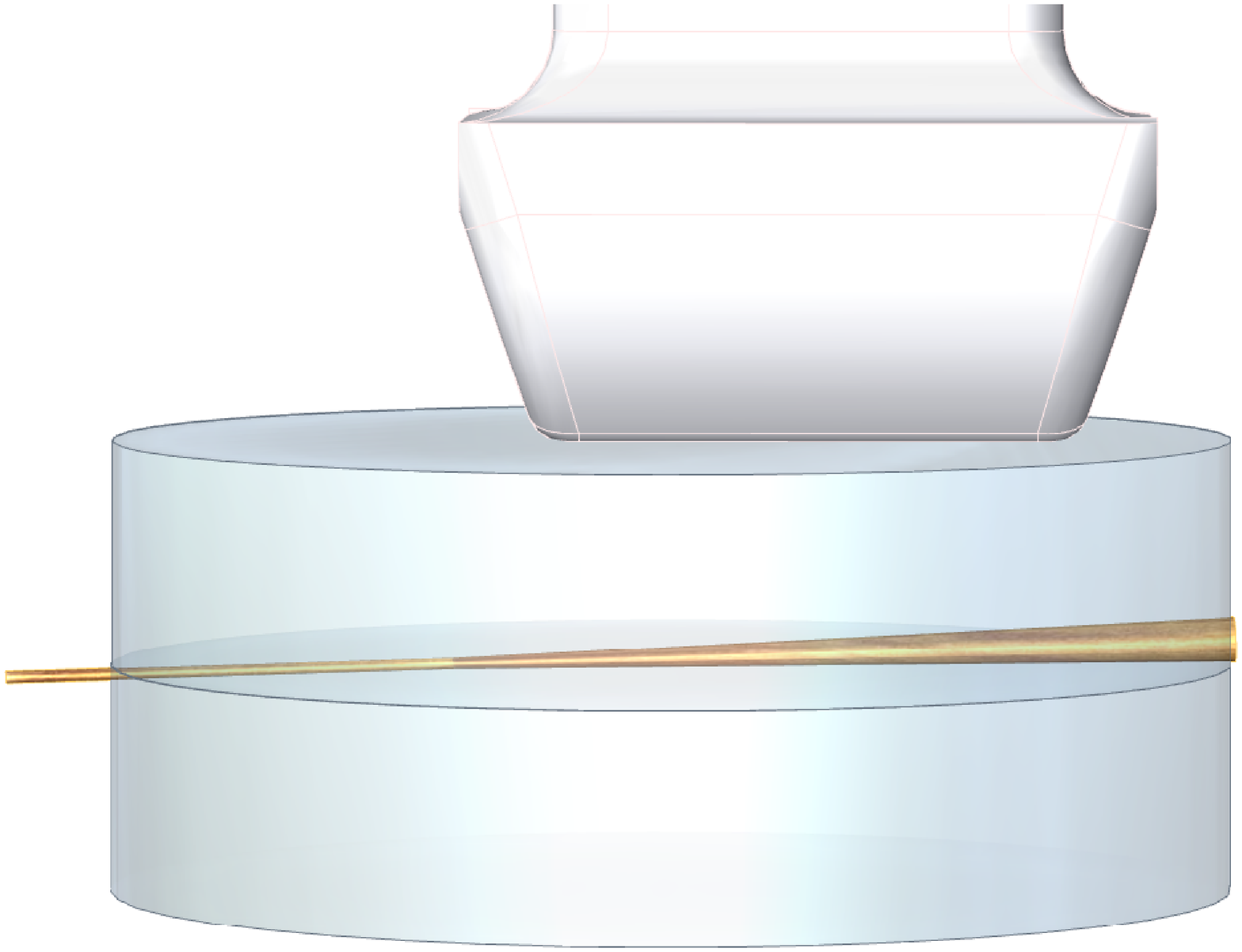}}}
\put(0.53,0.01){\reflectbox{\includegraphics[trim= 0 0 0 0,clip,width=0.2\unitlength]{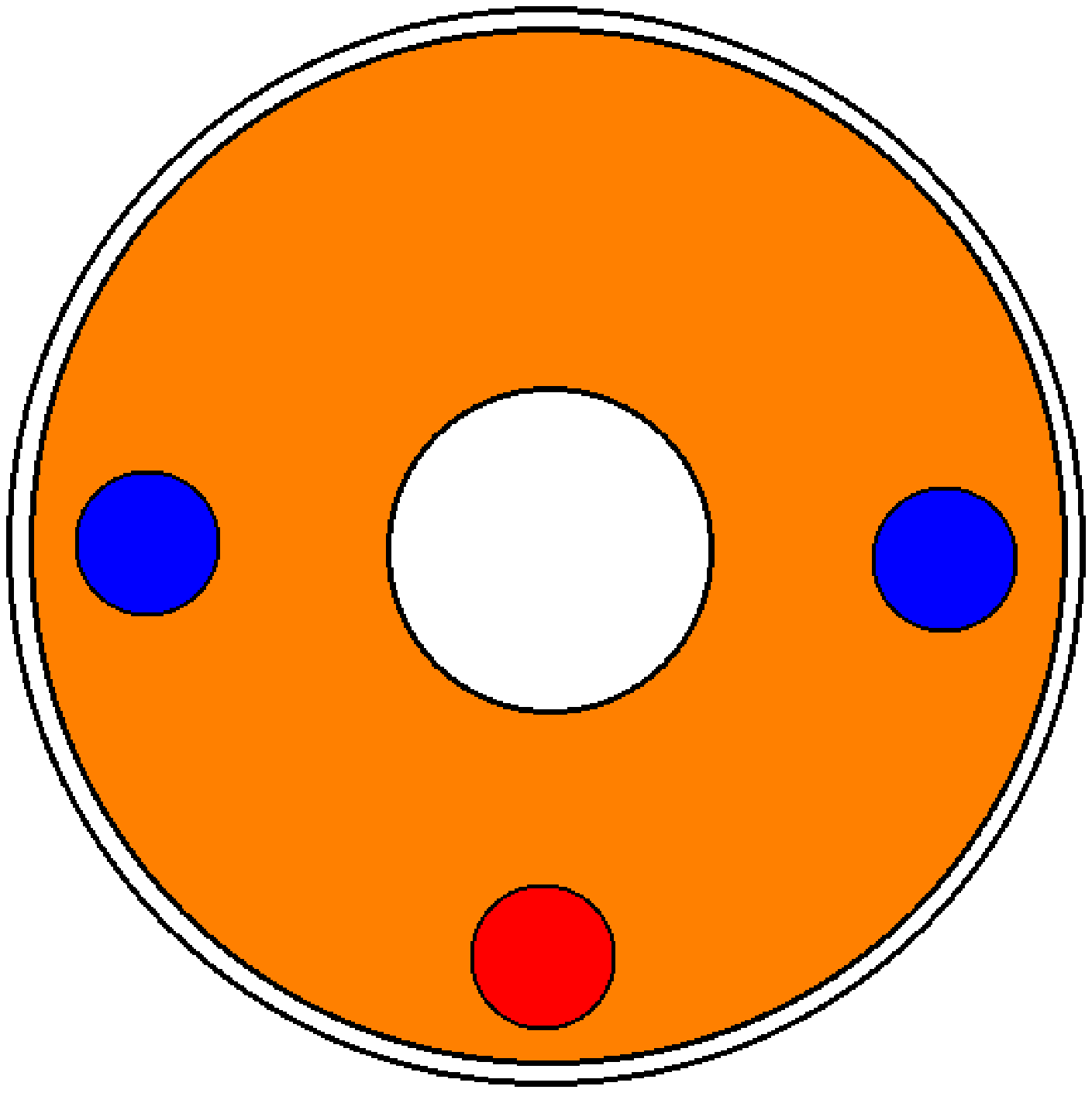}}}

\multiput(0.08,0)(0, 0.4){2}{\line(1,0){0.3}}
\multiput(0.08,0)(0.3, 0){2}{\line(0,1){0.4}}

\put(0,0.057){\scriptsize Syringe}
\put(0,0.04){\scriptsize pump}
\put(0.17,0.42){\scriptsize Transducer}
\put(0.17,0.3){\scriptsize Water bath}
\put(0.16,0.165){\scriptsize Flow direction}
\put(0.28,0.19){\vector(-1,0){0.1}}
\put(0.385,0.21){\scriptsize Microtube}
%\put(0.52,0.19){\vector(-1,0){0.07}}
\put(0.38,0.09){\scriptsize PBS + MBs}
\put(0.64,0.43){\scriptsize Transducer}
\put(0.8,0.36){\scriptsize Gelatin pad}
\put(0.58,0.33){\scriptsize Mouse tail}
\put(0.58,0.25){\scriptsize Gelatin pad}

\put(0.75,0.19){\scriptsize Vertebra}
\put(0.745,0.19){\vector(-2,-1){0.12}}
\put(0.75,0.16){\scriptsize Lateral Veins}
\put(0.745,0.165){\vector(-3,-1){0.18}}
\put(0.745,0.165){\vector(-1,-1){0.05}}
\put(0.75,0.03){\scriptsize Artery}
\put(0.745,0.035){\vector(-1,0){0.1}}

\put(0.03,0.44){\bf A}
\put(0.9,0.44){\bf B}
\put(0.85,0.05){\bf C}
\end{picture}
\end{flushright}
\caption{{\bf A} The microtube phantom experiment to quantitatively validate  super-resolution velocity estimation. A 200~$\mu$m diameter microtube was placed under an ATL L7-4 ultrasound probe. A reservoir of MBs was suctioned by a syringe pump at a constant flow rate. {\bf B} Experimental setup for the {\it in-vivo} mouse experiment illustrating the position of the mouse tail was placed between two gelatin pads and the ATL L7-4 ultrasound transducer. A volume of 100$\mu$L with a concentration of $2\times10^7$~MB/mL was injected into the mouse tail vein. {\bf C} Schematic of a cross section of a mouse tail illustrating the main vessel positions.}
\label{fig:methods}
\end{figure}

\subsection*{Calibrated sub-resolution flow phantom}

To verify that the proposed super-resolution velocity estimation method was able to correctly determine the flow rate and flow profile with sub-resolution accuracy, a calibrated flow phantom was fabricated (Figure \ref{fig:methods} {\bf A}). A polyethylene terephthalate glycol (PETG) microtube, with an internal diameter of 200~$\mu$m and a wall thickness of $20~\mu$m, was placed approximately 4~cm away, axially, from the transducer surface. For this depth a conventional B-mode elevational resolution is given by 1.6~mm.  Microbubbles were diluted in phosphate buffered solution (PBS) to a concentration of $2\times10^6$~MBs/mL~\cite{shelton2015}.  The mixture of PBS and MBs was placed in a cup and continuously stirred with a magnetic stirrer. A syringe connected to the other end of the microtube was mounted to a syringe pump (Harvard Apparatus, PHD 2000), which operated at calibrated flow rates, $Q$, ranging from 10 to 90~$\mu$L/min. The refill mode allow a continuous stirring of the PBS and MBs mixture ensuring a consistent concentration within the microtube throughout the experiment. The length of the microtube was 30 mm. At its extremes, a wider tube was attached (stick hermetically with hot glue) to give room to reach the pump and reservoir. Although, the microtube was intended to be aligned with the active phase of the transducer, a small angle was detected of approximately 0.5 degree. The system was immersed in degassed water to avoid external bubbles. Before proceeding to image acquisition, for every change of flow rate, the system was allowed to flow for 5~min to reach a steady state. The acquisition time per flow rate tested was 3s. In addition, every 20~min, the PBS-MBs solution was replaced by a new one, and the tubes were totally flushed, to minimize any changes in bubble concentration. For each imposed flow rate $Q$, a set of 2000 frames were acquired, at 700 frames per second, using our plane wave sequence.

\subsection*{\textit{In-vivo} super-resolution flow estimation in the mouse tail}
\label{sec:invivo}
{\it In-vivo} measurements of blood flow were performed in the mouse tail which was chosen due to its small size, accessibility to ultrasound and of its straightforward vascular structure compared to other potential targets such as the kidneys or tumors. Animal studies were approved by the University of North Carolina Institutional Animal Care and Use Committee. Six-week-old nude mice were anesthetized with isofluorane. During the experiment, the mice were kept warm with a heat lamp. The mouse tail was placed between two 2~cm thick gelatin pads (Aquaflex Ultrasound Gel Pad, Fairfield, NJ, USA) as illustrated in Fig. \ref{fig:methods} {\bf B}.  For the depth of 2~cm, a conventional B-mode elevational resolution would be 788~$\mu$m. By wetting the gelatin pads before tail placement a thin layer of degassed water was introduced between the pads and the tail. This layer of water helps to couple the gelatin pads with the tail and to avoid any air bubbles. A catheter, which was inserted towards the end of one of the tail veins, was used to inject a volume of 100~$\mu$L of PBS-MBs mixture with a concentration of $2\times10^7$~MB/mL. The injected concentration is higher than in the phantom experiment to compensate for dilution in blood circulating in the mouse so that the final concentrations in the vessels and microtube is approximately the same. The ultrasound probe was coupled with a layer of water on the top gelatin pad. A waiting period of 30 s was set after injecting the MBs and before starts the acquisition. The plane wave imaging sequence was used to acquire 6000 frames at 700 frames per second for each mouse tail experiment. The acquisition had a duration of approximately 9~s. The transducer was positioned by utilizing a live B-mode sequences and using as a guide the catheter.\\

\subsection*{Super-resolution signal processing and SVD filtering}

A number of processing stages are required to produce ultrasound super-resolution images and the velocity maps. First, an interpolation of the raw ultrasound data was performed to increase the native scanner sampling rate. Then, a delay-and-sum algorithm was applied to generate beamformed RF data. The RF data was then filtered with a singular value decomposition algorithm to reject tissue signal. This filtered RF data was used to spatially localize the microbubbles. Finally, the velocity was estimated by a cross-correlation algorithm applied to the microbubbles. These stages are, in general, common to most super-resolution approaches, however their specific implementation have a significant impact on the quality of the final images. Details of the specific implementation performed here are described below in the following subsections.

Data acquisition was performed at 4 times the center frequency, $F_s=20.83$~MHz.  To reduce the time-of-flight beamforming error and to improve the tracking accuracy \cite{pinton2006rapid} the raw ultrasound data was interpolated by a factor of four using cubic spline interpolation. This increased the sampling frequency to $F_s = 83.32$~MHz. Then, conventional delay-and-sum was used to generate a stack of beamformed RF data i.e. an RF movie of the microtube or tail. The field of view ranged from  16 to 25 mm laterally and from 2-4~mm in depth depending on the specific experiment, and processing was performed in this region of interest. The discretization of the image was set to an uniform pixel size of 29.8~$\mu$m, which is equivalent to $\lambda/10$, in both the depth and lateral dimensions. This is consistent with a recent study has found that an optimal grid discretization that minimizes the quantization error is $\lambda /2$ along with a upsampling, by means of an interpolation scheme, up to $\lambda/20$  \cite{song2018}. Thus, our beamforming grid was chosen to be very fine ($\lambda/10$) but no post-beamforming interpolation was performed.

The high frame-rate RF captures motion in two different spatio-temporal scales.  The first varies slowly with the time and uniformly in space and it is associated with bulk tissue motion and/or with the stationary background signal. The second is produced by the moving contrast agents that are spatially localized and are flowing within the vasculature or microtube. In our particular case, bulk tissue motion is minimum due to the anatomical region of interest (mouse tail). However, the nearly stationary background signal is still present. Furthermore, typically the MB signal is 20 to 30~dB lower than the background tissue signal. It is therefore  necessary to separate these two types of signal to visualize the MBs. In the literature, there are two main tools to separate the MBs signal from the tissue signal. The first is nonlinear imaging, in which one takes advantage from the fact that the MBs behaves much more nonlinearly than the surrounding tissue. Thus, techniques such as pulse inversion are useful to obtain MBs signal \cite{viessmann2013acoustic,oreilly2013superresolution}. However, this type of imaging provides limited improvement {\it in-vivo} for techniques that are not optimized, for instance, compensation in nonlinear response of the transducers itself. The second approach to separate the MBs signal from the tissue is the use of the singular value decomposition (SVD) filter.  This filter exploits the difference in coherence between the two types of signals. Tissue pixels are mostly coherent over a big area of the image, while the MB signal has a much shorter coherence length.  SVD filtering was first proposed to be as a clutter filter ~\cite{Ledoux1997,Bjaerum2002}. Although it has also been used in Doppler and functional ultrasound imaging~\cite{demene2015spatiotemporal}. Recently, SVD filtering has become a standard technique to extract the MB signals for super-resolution imaging \cite{errico2015ultrafast,lin2016super,lin20173-D,espindola2017adaptive,espindola2018adaptive}. 

Here, the cut-off values of the SVD filters were chosen empirically. A sweep in the cut-off of the filter was performed from lowest to highest cutoff until a signal was visible microtube/tail region. For the microtube phantom experiment, the highest five (out of 2000) singular values were filtered out. For the {\it in-vivo} experiment, the data set was split into two subsets of 3000 frames each and then the highest 10 (out of 3000) singular values were removed from each subset. The splitting of the data into two subsets was done to reduce the computational cost. The output of the SVD filter is a movie of the MBs that preserves motion along the microtube or vessels while suppressing the background signal which can be clutter and/or tissue. These filtered MB signals appear as moving point spread functions (PSFs) with a size that is consistent with the diffraction-limited resolution of the imaging system.

\subsection*{Microbubble localization}
The step after the SVD filtering consists of localizing the MB positions. Several methods have been used for this purpose, such as onset detection, centroid fitting, Gaussian fitting, and others which have been previously compared~\cite{Christensen2017}. In this study, the centroid method was used due to its adequate computational performance and its ease of implementation. However, before applying the centroid method, a two dimensional moving average filter with a  10$\times$10 pixel ($\lambda\times\lambda$) size was used to remove peaks that are smaller than the PSF. Finally, the MB positions were detected in each frame and these positions were accumulated to produce the final super-resolved images.

\subsection*{Velocity estimation}
The proposed method estimates the velocity with a normalized cross-correlation algorithm and a mask that acts on the bubble positions at each frame to only keep displacement information from the bubbles and not from the background or noise. This mask is a simple binary matrix, of the same size as an ultrasound image frame, with assigned values of one at the MBs positions.

The continuous cross-correlation $c(x)$ for two signals $S_i(x)$, can be defined as:

\begin{equation}
    c(x)=\frac{\int_{-K/2}^{K/2} S_i(X)S_{i+1}(x + X)dX}{\sqrt{\int_{-K/2}^{K/2} S_i^2(X)dX \int_{-K/2}^{K/2} S_{i+1}^2(x + X)dX}}.
\end{equation}
\noindent   where the subscript $i$ represent the frame and the independent variable $x$ is the spatial location.
This function correlates portions of the two signal defined by the kernel size $K$. The maximum value of the correlation $c(x_{\mathrm{max}})$, within a given search window, defined by $r$, is used to determine  the local velocity $v_{i+1/2}$  for this portion of signal, defined as $v_{i+1/2}=x_{\mathrm{max}}F_s$. Where $F_s$ is the frame rate. 

The interframe cross-correlation tracking algorithm was applied to the SVD filtered images. In these images the signal-to-noise ratio (SNR) is large only where there is MB motion. A kernel size $K=4\lambda$ with a search window of $r=\lambda/5$
was used to estimate MB motion between subsequent frames. The kernel size was chosen to be consistent with the observation that the bubbles, at the chosen concentration, had a typical separation distance greater than $4\lambda$. Thus, with this kernel size there is an insignificant overlap between bubbles and, therefore the interframe cross-correlation encode the motion of individual bubbles.  Finally, the search window size was based on the maximum expected theoretical interframe displacement  for the microtube and the flow rates imposed. Here a single search windows size was used for all the flow-rates studied. This parameter could be adjusted to be larger for higher flow rates or smaller for lower rates. 

The output of the correlation algorithm are two-dimensional maps of the inter-frame displacement and the maximum correlation coefficient for each ultrasound frame. Due to the low concentration of MBs, low numbers are detected within each frame (typically 20~MBs/frame). Therefore, the interframe displacement maps contain very localized bubble displacement information and noise where bubbles are not present. Thus, the bubble motion was extracted by masking the inter-frame displacement map with the position of the previously localized MBs. In addition, a quality filter based on the maximum correlation coefficient was applied and only estimates with correlation coefficient over 0.9 were preserved. This operation removes false bubbles, regions of the signal that do not have a similar pattern between two frames, or bubbles that disappear from the field of view. 

To generate a map of the velocity distribution within a vessel or microtube, individual velocity estimates from bubble motion were averaged in time, or equivalently frames, to produce a single spatially super-resolved velocity map as a function of space. At some spatial locations, a low number of bubble events were detected due to statistical variability. Consequently, the velocity estimates at these locations are more sensitive to noise since the average velocity in time is calculated over a small number of detected MBs. An additional filter was therefore implemented to exclude velocity estimates that have less than 3 MB detection events. This removes approximately a 14\% of the localization events. A longer acquisition time would lower this percentage.

\subsection*{Theoretical model of velocity distribution and flow}

\begin{figure}
 \begin{flushright}
 \setlength{\unitlength}{0.8\textwidth}
\begin{picture}(1.03,0.42)(0,0)
%%%% frame
%\multiput(0,0)(0, 0.42){2}{\line(1,0){1.03}}
%\multiput(0,0)(1.03, 0){2}{\line(0,1){0.42}}
%%%%%%%%%
\put(0,0){\includegraphics[trim= 90 280 160 150,clip,width=0.6\unitlength]{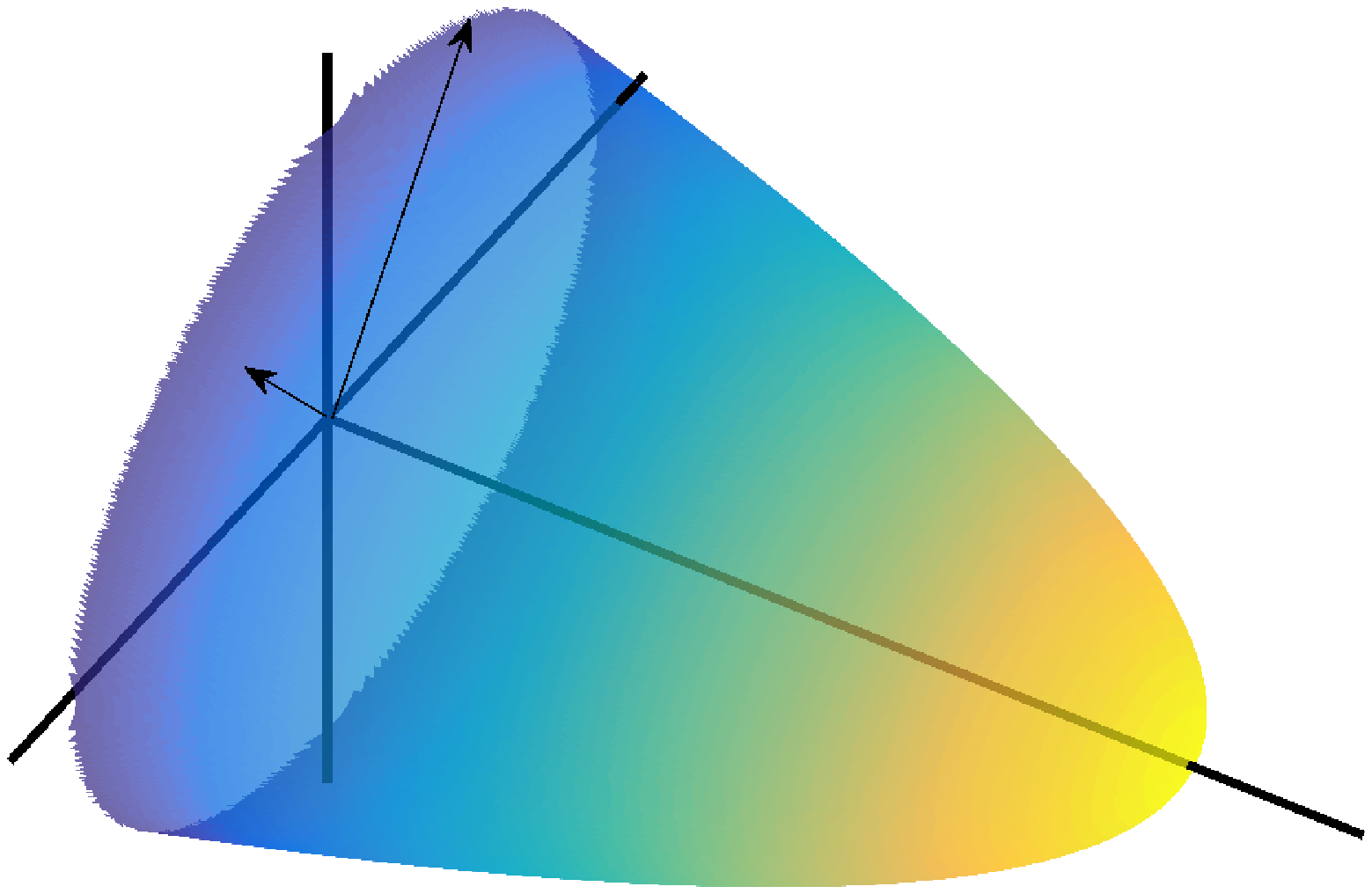}}
\put(0,0.1){$y$}
\put(0.12,0.36){$x$}
\put(0.2,0.36){$R$}
\put(0.1,0.205){$r$}
\put(0.57,0.01){$v(r)$}
\put(0.505,0.075){$v_{\mathrm{max}}$}
\put(0.6,0.1){\includegraphics[trim= 50 40 50 30,clip,width=0.4\unitlength]{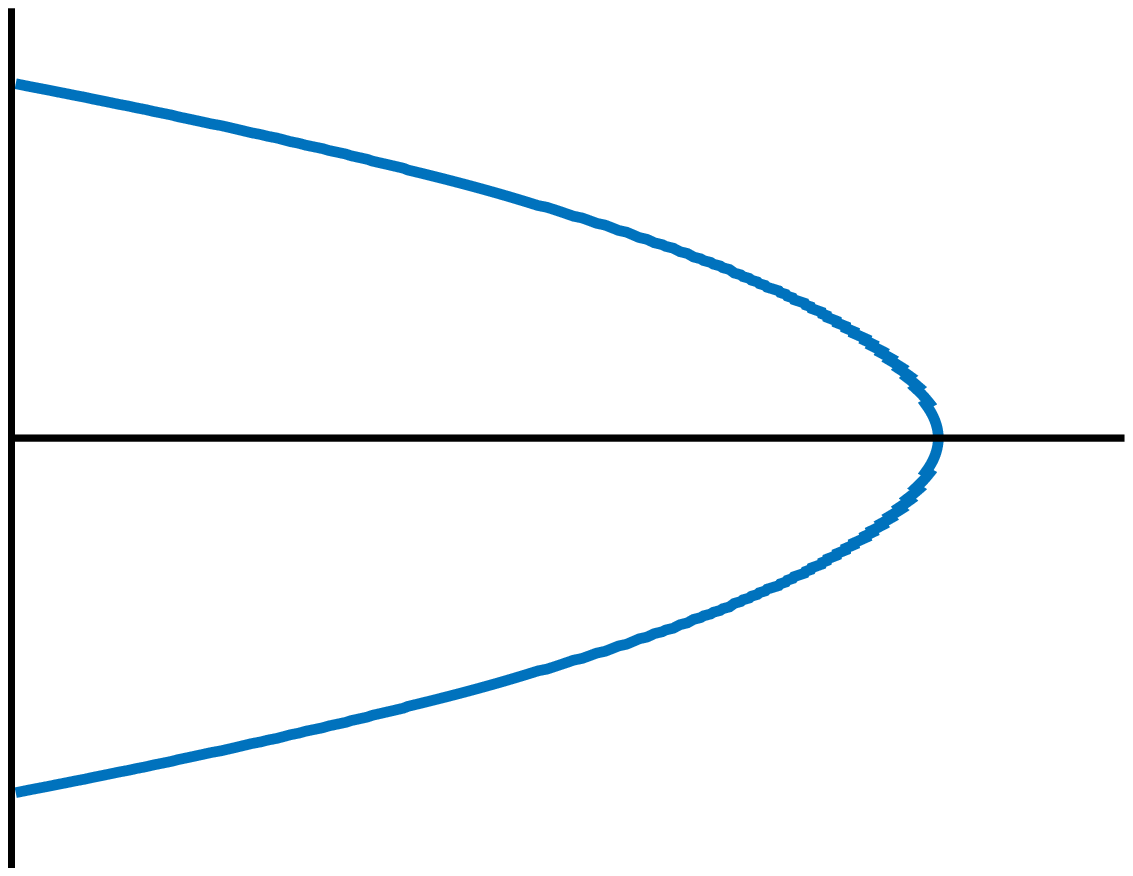}}
\put(0.61,0.1){$x$}
\put(0.97,0.22){$v(x)$}
\put(0.94,0.27){$v_a$}
\put(0.57,0.115){$R$}
\put(0.54,0.37){$-R$}
\end{picture}
\end{flushright}
    \caption{Scheme for velocity measurements in a tube. Left, magnitude of the local velocity as a function of the distance from the center of the tube $r$, inside a microtube with radio $R$. Right, velocity profile after has been collapsed in the elevation plane ($y$-axis).}
    \label{fig:3dtube}
\end{figure}

A flow model was developed to compare the experimentally determined velocity estimates to theoretical predictions from fluid dynamics. This model establishes a relationship between velocity, which is estimated in a two-dimensional imaging plane and the flow, which occurs in three dimensions. Classical fluid dynamics predicts a parabolic velocity profile in a tube with constant cross-section \cite{landau1959course} with radius $R$ (Fig. \ref{fig:3dtube} A). The model assumes that the velocity field is orthogonal to the microtube cross-section and that the velocity is zero at the walls. Our model also assumes a uniform weighting function within the elevation plane, which implies vessels with a diameter that is smaller that the elevational resolution. Then the maximum velocity, $v_{\mathrm{max}}$, at the center of the tube's cross section can be related to the velocity distribution in polar coordinates, with a quadratic equation:

\begin{equation}
 v(r) = v_{\mathrm{max}}(1-r^2/R^2)
 \label{eq:vmax}
\end{equation}
where $r$ is the radial coordinate from the center of the tube's section.

Although the imaging plane is two dimensional, physically the imaging process integrates information from the third (elevation) dimension. For a sub-resolution microtube or vessel that is oriented within the imaging plane, it is assumed that the velocity profile is integrated along the elevation dimension. This is consistent with the point spread function shape in the elevation dimension being much larger than the microtube/vessel width. For example here the microtube diameter is 200~$\mu$m and the elevation resolution at the tube depth is 1.6~mm. Therefore the velocity measurement shown in the imaging plane represents an average of the parabolic velocity profile along the elevation dimension, $y$. This can be expressed as,

\begin{equation}
 v(x) = \frac{\int_{0}^{\sqrt{R^2-x^2}}v(x,y)dy}{\sqrt{R^2-x^2}} 
 \label{eq:vx}
\end{equation}

\noindent where $v(x,y)$ is $v(r)$ expressed in Cartesian coordinates. Then, by substituting the velocity profile in Eq.~\ref{eq:vmax}, this integral can be evaluated as:

\begin{equation}
v(x) = \frac{2}{3}v_{\mathrm{max}}(1-x^2/R^2)
 \label{eq:vx2}
\end{equation}

\noindent Fig. \ref{fig:3dtube} B shows the integrated version of the three-dimensional velocity profile described by Eq.~\ref{eq:vx2}. This equation also shows that the peak velocity measured in the 2D imaging plane $v_a$ is 2/3 as large as the peak velocity in 3D, i.e. 

\begin{equation}
        v_{a}=2/3v_{\mathrm{max}}
    \label{eq:va}
\end{equation}

\noindent Then, conservation of mass can be used to relate $v_{\mathrm{max}}$ to the flow rate, $Q$. For a microtube with a cross sectional area $S$, the flow rate is:

\begin{equation}
 Q=\int_S \vec{v}(\vec{r})\cdot d\vec{A} = \frac{\pi}{2}R^2v_{\mathrm{max}}
\end{equation}

\noindent  where $d\vec{A}$ is a differential element of area. This last equation can be  re-expressed as the maximum velocity within the tube, a quantity that is measured by the proposed imaging method, as a function of the flow rate, a known quantity determined by the syringe pump:

\begin{equation}
v_{\mathrm{max}}=\frac{2Q}{\pi R^2}
\label{eq:v_q}
\end{equation}

\noindent Finally, using Eq. \ref{eq:vmax} and \ref{eq:v_q} it is possible to relate the imposed flow $Q$ of the microtube/vessel with the maximum velocity at the center of the flow $v_{\mathrm{max}}$ and with the expected experimental measurement of the maximum velocity $v_a$ after the integration over the elevation direction.

\section*{Results}

\subsection*{Microtube phantom validation}

The calibrated microtube flow phantom was used to experimentally validate the theoretical flow model. The proposed velocity estimation method was applied to the microtube phantom data and then scaled by the elevation plane integration factor to compare the imposed flow with the experimentally obtained maximum velocities.

\begin{figure}[t]
\begin{flushright}
\setlength{\unitlength}{0.84\textwidth}
\includegraphics[width=\unitlength]{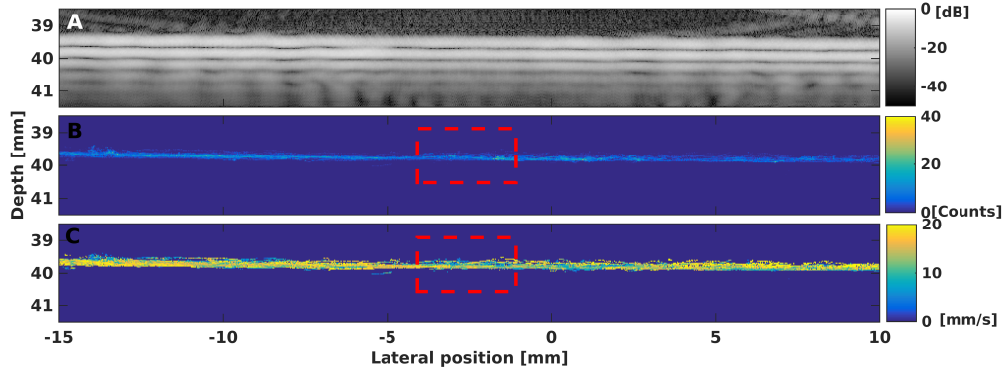}
\end{flushright}
\caption{B-Mode image ({\bf A}), super-resolved image ({\bf B}) and velocity map ({\bf C}) of a 200~$\mu$m diameter microtube with a flow rate of 30$\mu$L/min. The velocity scale is slightly saturated to allow a better visualization. The actual range of the scale is of 29 counts. The rectangular boxes indicate the portions of the images used to calculate flow profiles.}
\label{fig:tube_images}
\end{figure}

The B-mode image  (Fig. \ref{fig:tube_images}{\bf A} shows that with this conventional imaging method the microtube appears to be larger than its actual size. This is due to the diffraction limit of resolution, and multiple reverberation which can be seen as a repetition of the image over periodic intervals in depth. Figure \ref{fig:tube_images}{\bf B} shows the super-resolved images generated from the accumulation of all the positions of the detected MBs within all the 2000 frames. The color scale in the image, labeled count, represents the number of bubbles detected at each pixel. A total of 40979 bubbles were detected, which is, in this case, sufficient to clearly visualize the microtube across the 25 mm field of view. This corresponds to an average of 20.5 MBs per frame. Then, a lateral sum of the super-resolved image within the dashed rectangle was used to determine the tube profile (Fig. \ref{fig:tube_profile} A). The width of this profile, measured at the full width at half of the maximum (FWHM), is 170$\pm$50~$\mu$m, which is consistent with the 200~$\mu$m internal diameter reported by the microtube manufacturer. Note that this number was determined from the average of all the experiments performed for different imposed flows, and the error is given by the corresponding standard deviation.

\begin{figure}[t]
\begin{flushright}
\setlength{\unitlength}{0.84\textwidth}
\includegraphics[width=\unitlength]{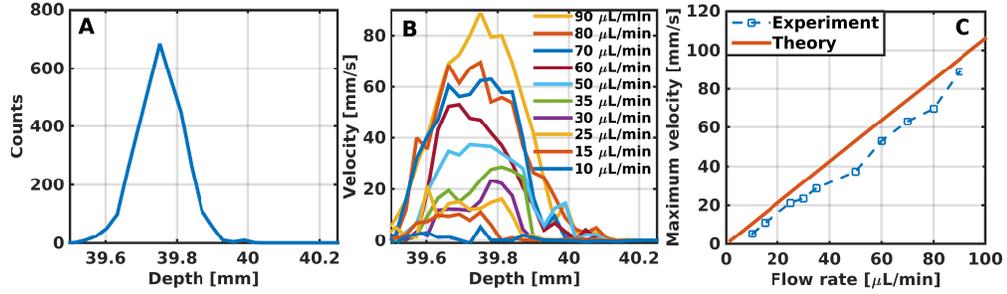}
\end{flushright}
\caption{{\bf A} Vertical profile of the super resolved images shown in Fig. \ref{fig:tube_images}{\bf B} averaged between -4.2~mm and -1.2~mm laterally. {\bf B} Velocity profiles obtained from the velocity (also an average from -4.2~mm and -1.2~mm laterally). C Comparison between the known velocity calculated with Eq. \ref{eq:v_q} (red line) and measured experimentally from the peak velocity in Fig. \ref{fig:tube_profile}{\bf B} (dashed blue line) as a function of the imposed flow rate. The results shown here includes the correction factor} 
\label{fig:tube_profile}
\end{figure}

For the particular case shown in Fig. \ref{fig:tube_images}, where the flow rate was set to $30~\mu$L/min, the proposed technique measured bubble velocities in a range between 1 and 33~mm/s. The average velocity, calculated from all pixels with a non-zero velocity, was 14.37~mm/s. The velocity profile, calculated as a lateral average within the rectangle (The observe angle of 0.5 degree was corrected by rotating the image anticlockwise), and scaled by a factor of 3/2 following the scaling factor determined in Eq. \ref{eq:vmax}, yields the physical velocity estimate of for the case of the phantom experiments, only the maximum velocity. This spatial average was necessary to reduce the noise in the velocity estimation, however,  2000 frames (less than 3s of acquisition) were acquired per each flow rate imposed. This low number of frames was set to reduce the total duration of the experiment, given the waiting time to stabilize the flow and given the 20 minutes duration of the bubbles once diluted. Thus, as the number of frames is increased, the necessity for a spatial average is reduced. These velocity profiles were calculated for flow rates ranging between 10$\mu$L/min to 90$\mu$L/min and they are shown in Fig. \ref{fig:tube_profile}{\bf B}. In these set of experiments the proposed velocity estimation technique detected local velocities as small as 1~mm/s;

The maximum velocities, $v_{\mathrm{max}}$, for each flow rate experiment, determined from the physical velocity profiles in Fig. \ref{fig:tube_profile}{\bf B}, are shown in Fig. \ref{fig:tube_profile}{\bf C} 
(dashed blue line). These measured maximum velocities can be compared to theoretical prediction based on the known flow rate in the microtube using Eq. \ref{eq:v_q} (Fig. \ref{fig:tube_profile} {\bf C}, red curve). The maximum velocity as measured by the proposed imaging method depends linearly on the imposed flow rate, as expected. Furthermore, it matches theoretical predictions. There is, however a small but consistent underestimation of the experimental maximum velocity of 9~mm/s in average. This underestimation produces a 49\% error at the smallest flow rate (10~$\mu$L/min) tested but a 7\% error for the highest flow rate (90~$\mu$L/min). For low velocity of the bubbles, the correlation algorithm produces high percentage error due to low axial resolution which is comparable to the interframe bubble motion. At high flow rates, this error is reduced due to the increased interframe displacement of individual bubbles. In addition, we speculate that this error is originated from the use of the cross-correlation algorithm, which integrates the velocity within the kernel window.

\subsection*{Blood velocity in the {\it in-vivo} mouse}

After the quantitative validation of the proposed technique, experiments were performed to demonstrate the feasibility in {\it in-vivo} configurations with two mouse tails. A clinically relevant microbubble concentration ($2\times10^7$MB/$\mu$ L)  was injected into the mouse tail with the help of a catheter. This experimental procedure was performed as described previously in section \ref{sec:invivo}. With a  conventional B-mode image (Fig. \ref{fig:mouse2}, A) the mouse tail is not distinguishable from the background speckle. This is probably due to the size of the anatomical structures which are significantly smaller than the imaging wavelength. Furthermore, the weak reflections from small structures in the tail, such as a vertebra, gelatin pad walls, and skin, can be easily overcome by sources of image degradation such as clutter and reverberation \cite{Pinton211Sources}. However, in the super-resolved image (Fig. \ref{fig:mouse2}, B) the mouse tail vein is clearly visible. No significant angle between the transducer phase and the vein was found. The total number of MBs detected was 85142 within 6000 frames, which corresponds to an average of 14 MBs per frame. The corresponding velocity map is shown in Fig. \ref{fig:mouse2}{\bf C}. As expected for a vein, this map shows a flow toward the head of the mouse. 

\begin{figure}[t]
\begin{flushright}
\setlength{\unitlength}{0.84\textwidth}
\includegraphics[trim=0 0 0 0,clip,width=\unitlength]{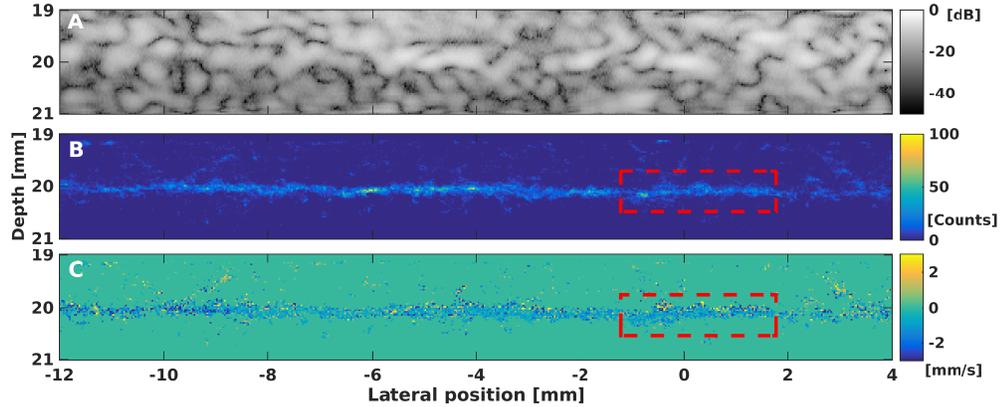}
\end{flushright}
\caption{Images from the second mouse tail experiment. {\bf A} Conventional B-mode image, {\bf B} Super-resolved image. {\bf C} Color coded velocity map, yellow indicates a velocity vector that point towards the end of the tail while blue indicates a velocity vector towards the head. The rectangular boxes indicate the portions of the images used to determine flow profiles.}
\label{fig:mouse2}
\end{figure}

The mouse vein profile (Fig. \ref{fig:mouse2prof}{\bf A}) was determined by averaging the MBs within the red box drawn in Fig. \ref{fig:mouse2}{\bf A}. The diameter of the vein was estimated to be 140~$\mu$m at the FWHM. The average velocity within the region defined by the box produces the velocity profile shown in Fig. \ref{fig:mouse2prof} {\bf B}. This profile has a characteristic parabolic shape and a peak velocity of about 0.8~mm/s. Note that this velocity profile has not been corrected by the spatial integration in the elevation direction. Therefore, assuming the vein is aligned within the imaging plane, the real physical peak velocity should be in the order 1.2 mm/s. Thus, by using Eq. \ref{eq:v_q} a flow rate of approximately 0.6$\mu$L/min can be estimated.

\begin{figure}[t]
\begin{flushright}
\setlength{\unitlength}{0.84\textwidth}
\includegraphics[width=\unitlength]{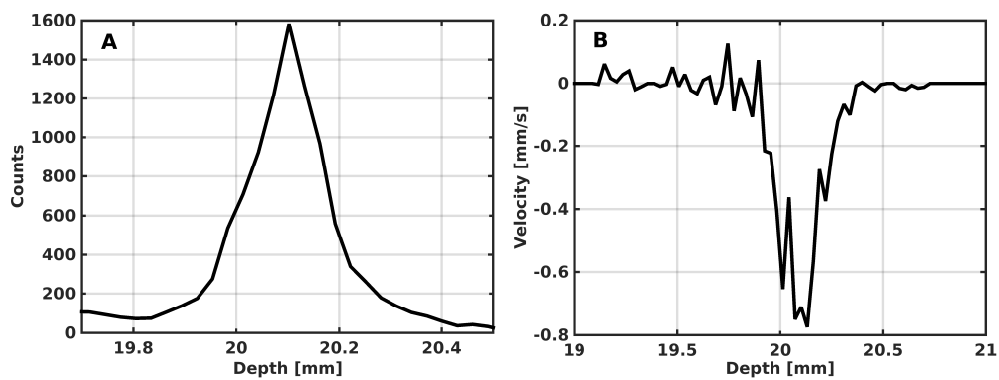}
\end{flushright}
\caption{{\bf A} Super-resolved profile computed from the accumulation of count within the box shown in Fig.\ref{fig:mouse2}{\bf B}. {\bf B} Average velocity profile inside the box shown in Fig. \ref{fig:mouse2}{\bf C}.}
\label{fig:mouse2prof}
\end{figure}

\begin{figure}[ht]
\begin{flushright}
\setlength{\unitlength}{0.84\textwidth}
\includegraphics[width=\unitlength]{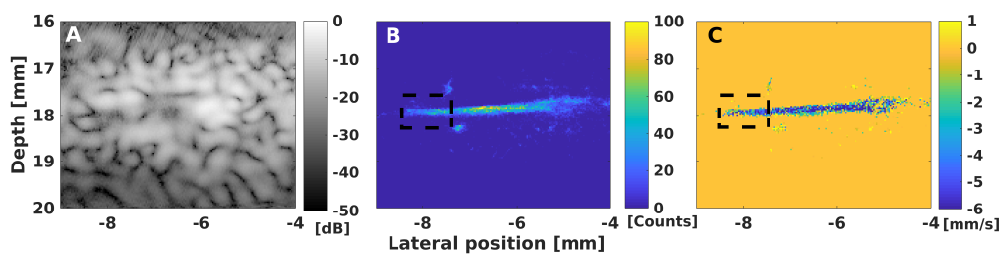}
\end{flushright}
\caption{Image from the first mouse tail experiment. B-mode image {\bf A}, super-resolved images {\bf B} and velocity map {\bf C}. In the color coded velocity map, yellow indicates a velocity vector that point towards the end of the tail while blue indicates a velocity vector towards the head. The rectangular boxes indicate the potions of the images used to trace the respective profiles.}
\label{fig:mouse1}
\end{figure}

In the second mouse tail experiment, the B-mode image, Fig. \ref{fig:mouse2}{\bf A} looks similar to the image of the first mouse, i.e. the tail is not clearly visible. The super-resolved image (Fig. \ref{fig:mouse1}{\bf B}), however, clearly shows the presence of a sub-resolution vessel with an angle of 2.9 degree with respect to the transducer array line. The flow within the vessel can be seen in the velocity map (Fig. \ref{fig:mouse1} {\bf C}). Here negative values represent velocity to the left, which is towards the mouse head, indicating that the vessel is a tail vein. The spatially super-resolved velocity images of the mouse tail show a vein visible along a  4~mm extent of the imaging field of view. 

\begin{figure}[ht]
\begin{flushright}
\setlength{\unitlength}{0.84\textwidth}
\includegraphics[width=\unitlength]{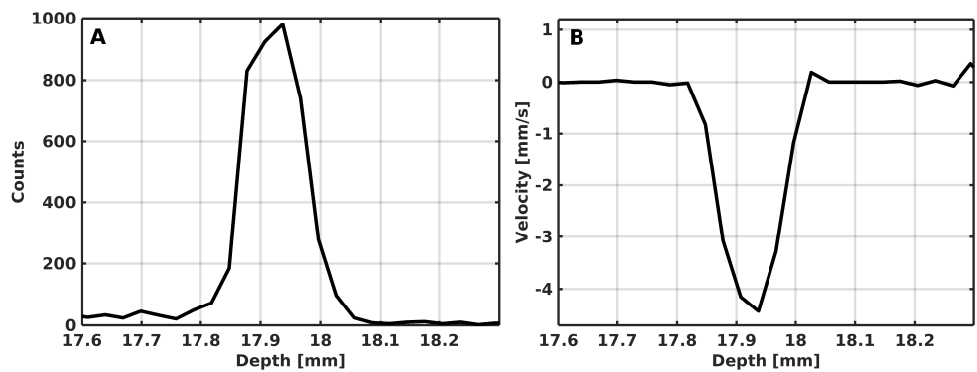}
\end{flushright}
\caption{{\bf A} super resolved profile of the vein calculated by summing laterally within the black box shown in Fig. \ref{fig:mouse1}{\bf B}. {\bf b} Velocity profile of the vein obtained by averaging the velocity in the same boxed region as the velocity map shown in Fig. \ref{fig:mouse1}{\bf C}.}
\label{fig:mouse1Prof}
\end{figure}

The total number of detected MBs was 101994 within the 6000 frames that were acquired. This corresponds to an average of 17 MBs per frame. By summing these bubble detection events laterally within the black box, the vein profile can be calculated (Fig. \ref{fig:mouse1Prof}{\bf A}). The size of the vein, calculated as the FWHM of this profile, was found to be 120~$\mu$m. A lateral average of the velocity map within the black box yields the velocity profile of the vein (Fig. \ref{fig:mouse1Prof}). The velocity profile of the vein has a nearly parabolic shape with a peak velocity of -4.4~mm/s.

\section*{Discussion}
\label{discussion} 

In the present paper, a configuration to measure one-dimensional flow was chosen for simplicity. Two-dimensional measurement of the flow would be possible just by computing the two-dimensional cross correlation. However, with the proposed technique three-dimensional velocity measurements are not possible since our ultrasound probe is only sensitive to in-plane motion. Thus, any out-of-plane motion is filtered out, underestimating the measurement of the velocity of the flow. Therefore, a matrix ultrasound array could improve the velocity measurements by its potential ability to measure both, the in-plane and out-of-plane velocities. The matrix ultrasound probe also would improve the spatial integration problem by making the elevation axis indistinguishable from the lateral axis.

The mouse tail contains shallow vascular structures. This was the reason behind the election of a plane wave imaging sequence. However, we think that this method could also be applied using focused sequence, such as the one presented in \cite{espindola2017adaptive,espindola2018adaptive}, to reach information of deep structures.

In the previous calculation that ends with Eq. \ref{eq:v_q}, the velocity vector was restricted to the imaging plane, i.e., no out of plane angle was considered. In general, non zero angles $\theta$ that are outside of the imaging, i.e. in the elevation dimension, might exist. This would produce an out-of-plane component of the velocity. In this case, equation 5 would incorporate a factor $\cos(\theta)$ which would reduce further the estimated of the velocity when compared to the real maximum velocity, i.e,
 \begin{equation}
     v_{\mathrm{max}} = \frac{3}{2\cos(\theta)}v_a
 \end{equation}
 Thus, the correction 3/2 acts independently of the angle when the velocity field is projected over the corresponding imaging plane. However, different correction factor might be expected in cases where the section of the flow deviates from a circular shape. This is a standard calculation/limitation for Doppler techniques \cite{jensen1996estimation}.
 
To compute the velocity profiles an spatial average was necessary to reduce the noise in the velocity estimation, however,  only 2000 frames (less than 3s of acquisition) were acquired in each experiment. This low number of frames was set to reduce the total duration of the experiment, given the waiting time to stabilize the flow and given the 20 minutes duration of the bubbles once diluted. Thus, as the number of frames is increased, the necessity for a spatial average is reduced.

For vary small vessels where the blood act more like a suspension of blood with plasma, the flow as been reported to behave in a more complex way \cite{lanotte2016red}. In this case the proposed correction of the flow would be invalid.

\section*{Summary and conclusion}

The proposed super-resolution velocity estimation technique provides quantitative velocity imaging of the vasculature within the {\it in-vivo} mouse tail along with parabolic velocity profiles of a vein. Even though the vessel diameter (140 $\mu$m) is smaller than the diffraction resolution limit ($1.22\lambda$Fnumber= 363 $\mu$m) this technique successfully resolves, not only the vessel, but the flow dynamics within this small vessel. A calibrated microtube phantom experiment demonstrated that the proposed super-resolution velocity estimation technique has a an error of 20\% when compared to theoretical predictions based on fluid dynamics and the thickness of the imaging plane. Furthermore, this comparison demonstrates that to accurately estimate the velocity inside a microvessel, the integration over the elevation direction must be accounted for. This consists of a simple multiplicative correction factor of 3/2 that scales the  measured peak velocity to obtain the physical peak velocity within the vessel. This correction is useful even in an out-of-plane case. As Eq. 8 indicates, this correction applied for the in-plane components of the velocity. The use of volumetric imaging with an ultrasound matrix array that could image the third velocity component in the elevation dimension could reduce or remove the need for a correction factor depending on the elevation plane thickness.  

The two mouse tail experiments suggest that the proposed method has a sensitivity of less than 1~mm/s. The first mouse tail experiment showed a peak velocity of the blood flowing through the tail vein as small as 1~mm/s, which is near the lower bound of what our technique can estimate. However, the second mouse tail experiment presented a peak velocity approximately four times bigger. At this relatively bigger flow velocity, our technique estimated a more reliable velocity profile which can be noted by its nearly parabolic shape.

\ack
This work was supported by grants from the National Institute of Health R01-CA220681, R01-EB025419.

%\References
\section*{References}
\bibliographystyle{dcu}
\bibliography{citation.bib}

\end{indented}

\end{document}